\documentclass{article}
\usepackage{spconf}

\usepackage{fullpage}
\usepackage{graphicx,float,amsmath,amsfonts,amssymb,xcolor,enumerate,array,bm,amsthm,multirow,tabu,mathrsfs,mathtools,dsfont}
\usepackage{url}
\graphicspath{ {../figures/}{./} }

\usepackage{nicefrac}
\usepackage{booktabs}

\newcommand{\half}{\nicefrac{1}{2}}

\newcommand{\ba}{{\rm\bf a}}
\newcommand{\bb}{{\rm\bf b}}
\newcommand{\bbeta}{\bm{\beta}}
\newcommand{\bh}{{\rm\bf h}}
\newcommand{\bx}{{\rm\bf x}}

\newcommand{\bp}{\bm{p}}
\newcommand{\br}{\bm{r}}
\newcommand{\bzero}{{\rm\bf 0}}

\newcommand{\GG}{{\rm\bf G}}

\newcommand{\TT}{{\rm\bf T}}
\newcommand{\RR}{{\rm\bf R}}

\newcommand{\la}{\left\langle}
\newcommand{\ra}{\right\rangle}
\newcommand{\e}{{\rm e}}
\newcommand{\di}{\,{\mathrm{d}}}
\newcommand{\dB}{{\rm dB}}

\newcommand{\transp}{^{\mbox{\tiny\sc H}}}
\newcommand{\transpr}{^{\mbox{\tiny\sc T}}}

\newcommand{\E}[1]{\mathbb{E}\left[ #1 \right]}

\newcommand{\C}{\mathbb{C}}
\newcommand{\R}{\mathbb{R}}
\newcommand{\bS}{\mathbb{S}}
\newcommand{\Z}{\mathbb{Z}}

\newcommand{\eqdef}{\stackrel{{\rm def}}{=}}

\mathtoolsset{showonlyrefs=true}%only show numbering of referenced equations

\def\hquad{\hskip0.5em\relax}
\DeclarePairedDelimiter\paren{\lparen}{\rparen}

\newlength{\Oldarrayrulewidth}
\makeatletter
\newcommand{\pushright}[1]{\ifmeasuring@#1\else\omit\hfill$\displaystyle#1$\fi\ignorespaces}
\newcommand{\pushleft}[1]{\ifmeasuring@#1\else\omit$\displaystyle#1$\hfill\fi\ignorespaces}
\makeatother

\title{FRIDA: FRI-BASED DOA ESTIMATION FOR ARBITRARY ARRAY LAYOUTS}
\name{Hanjie Pan$^\dag$, %
  Robin Scheibler$^\dag$, %
  Eric Bezzam$^\dag$, %
  Ivan Dokmani\'{c}$^\ddag$, %
  and Martin Vetterli$^\dag$ %
  \thanks{This work was supported by %
    the Swiss National Science Foundation grant % 200021\_138081 -- Non Linear Sampling Methods.
    20FP-1\_151073 --- Inverse problems regularized by sparsity. ID was funded by LABEX WIFI (Laboratory of Excellence within the French Program ``Investments for the Future'') under references ANR-10-LABX-24 and ANR-10-IDEX-0001-02 PSL* and by Agence Nationale de la Recherche under reference ANR-13-JS09-0001-01.
    All the code used to produce the results of this paper is available at http://go.epfl.ch/FRIDA.}}
\address{%School of Computer and Communication Sciences %
  $^\dag$\'{E}cole Polytechnique F\'{e}d\'{e}rale de Lausanne (EPFL), Switzerland \\%
  $^\ddag$Institut Langevin, CNRS, ESPCI Paris, PSL Research University, France \\%
  \{firstname.lastname\}@epfl.ch, ivan.dokmanic@espci.fr}

\ninept

\begin{document}

\maketitle

\begin{abstract}
In this paper we present FRIDA---an algorithm for estimating directions of arrival of multiple wideband sound sources. FRIDA combines multi-band information coherently and achieves state-of-the-art resolution at extremely low signal-to-noise ratios. It works for arbitrary array layouts, but unlike the various steered response power and subspace methods, it does not require a grid search. FRIDA leverages recent advances in sampling signals with a finite rate of innovation. It is based on the insight that for any array layout, the entries of the spatial covariance matrix can be linearly transformed into a uniformly sampled sum of sinusoids. 
\end{abstract}


%==============================================
\section{Introduction}
%==============================================

A wishlist for a direction of arrival (DOA) estimator may look something like this: it should be high-resolution, work at low signal-to-noise ratios (SNRs), resolve many possibly closely spaced sources, work with few arbitrarily laid out microphones, and do so efficiently, without grid searches. 

It is uncommon to have all of these items checked at once. For example, the steered response power (SRP) methods \cite{Capon:1969da} can be made robust, do not require a specific array geometry, and are immune to coherence in signals. Because they are based on beamforming though, they cannot resolve close sources \cite{DiBiase:2000uv}.

Close sources can be resolved by the high-resolution DOA finders. Their main representatives are subspace methods such as MUSIC \cite{Schmidt:1986js}, Prony-type methods such as root-MUSIC \cite{Friedlander:1993ew}, and methods that attempt to compute the maximum likelihood (ML) estimator such as IQML \cite{Bresler:1986hq}. 

Subspace methods exploit the fact that for uncorrelated signal and noise, the eigenspace of the spatial covariance matrix corresponding to largest eigenvalues is spanned by the source steering vectors \cite{Schmidt:1986js}. These methods are fundamentally narrowband since the signal subspaces vary with frequency; they can be made wideband either by incoherently combining narrowband estimates or, better, by combining them coherently through transforming the array manifold at each frequency to a manifold at a reference frequency (CSSM \cite{Wang:1985jv}, WAVES \cite{diClaudio:2001bb}). These methods require a search over space unless the array is a uniform linear array (ULA) \cite{barabell1983improving}. Coherent methods also require special ``focusing matrices'', essentially initial guesses of the source locations. WAVES can do without focusing but at the cost of performance. In between coherent and incoherent methods is the TOPS algorithm \cite{yoon2006tops}, which performs well at mid-SNRs, but still requires a search and performs worse than coherent methods at low SNRs.

We propose a new finite rate of innovation (FRI) sampling-based algorithm for DOA finding---FRIDA.
Among the mentioned algorithms, FRIDA is most reminiscent of IQML \cite{Bresler:1986hq}, especially for narrowband signals and ULAs. Unlike IQML, FRIDA works for arbitrary sensor geometries and for wideband signals. Moreover, it uses multi-band information coherently. Still, it requires no grid search and no sensitive preprocessing akin to focusing matrices, and it achieves very high resolution at very low SNRs, outperforming previous state-of-the-art. 

FRIDA can work with fewer microphones than sources as it uses cross-correlations instead of raw microphone streams. The tradeoff is that it is not able to handle completely correlated signals. A straightforward modification of the algorithm which operates on raw signals rather than cross-correlations does not have this issue, but it requires more microphones. 

The main ingredient of FRIDA is an FRI sampling algorithm \cite{vetterli2002sampling}. 
%Initially requiring uniform sampling grids and being rather sensitve to model mismatch and noise, 
FRI sampling has recently been extended to non-uniform grids along with a robust reconstruction algorithm~\cite{pan2016towards}. The algorithm is an iterative algorithm similar to IQML, but with an added spectral resampling layer and a modified stopping criterion (Section~\ref{sssec:reconAlg}).
The key insight is that the elements of the spatial correlation matrix can be \emph{linearly} transformed into uniformly sampled sums of sinusoids, regardless of the array geometry.

\section{Notation and Problem Formulation}

%Throughout the paper, matrices and vectors are denoted by bold upper and lower
%cases, respectively.  
%Let us first state the problem and establish the
%notation. The Euclidean norm of a vector $\br$ is denoted by 
%%$\|\br\| =
%%(\br^\top \br)^{\half}$. 
%We denote by $\bS$ the unit circle (unit sphere in 2D).
%Vectors denoted by $\bp$ are usually unit propagation vectors.
%
%\todo{Explicitely list all assumptions: far field, non-correlated, etc}
%
%In the source localization problem, we assume a setup with $Q$ microphones
%located at $\{\br_q\in\R^2\}_{q=0}^{Q-1}$, and $K$ monochromatic and
%uncorrelated, point sources in the far-field indexed by letter $k$.  Each
%propagates in the direction of the unit vector $\bp_k \in\bS$.  Within a narrow
%band centred around frequency $\omega$, the baseband representation of the
%signal coming from direction $\bp\in\bS$ is
%
%[ROBIN\_AND\_HANJIE]
Throughout the paper, matrices and vectors are denoted by bold upper and lower case letters. The Euclidean norm of a vector $\bx$ is denoted by $\|\bx\|_2 = (\bx\transp \bx)^{\half}$. We denote by $\bS$ the unit circle. Unit propagation vectors will be denoted by $\bp$.

%\todo{Explicitely list all assumptions: far field, non-correlated, etc}

\subsection{Source Signal and Measurements}

\subsubsection{Sources with Arbitrary Spatial Support}

We assume a setup with $Q$ microphones located at $\{\br_q\in\R^2\}_{q=1}^{Q}$, and $K$ monochromatic and uncorrelated point sources in the far-field indexed by the letter $k$.  Each propagates in the direction of the unit vector $\bp_k \in\bS$.  Within a narrow band centered at frequency $\omega$, the baseband representation of the signal coming from direction $\bp\in\bS$ reads
\(%\begin{equation}
x(\bp, \omega, t) = \tilde{x}(\bp, \omega) \e^{j\omega t},
\)%\end{equation}
where $\tilde{x}(\bp, \omega)$ is the emitted sound signal by a source located at $\bp$ and frequency $\omega$. The \emph{intensity} of the sound field is then
\begin{equation}
I(\bp, \omega)
\eqdef
\E{|x(\bp, \omega, t)|^2}
=\E{|\tilde{x}(\bp, \omega)|^2}.
\label{eq:mtx_xcorr}
\end{equation}
We assume frame-based processing, and the expectation is over the randomness of $\tilde{x}$ from frame to frame. As $\tilde{x}$ carries the phase, the assumption $\E{\tilde{x}} = 0$ holds. Note that at this point we did not yet make the point source assumption.

The received signal at the $q$-th microphone located at $\br_q$ is the integration of all plane waves along the unit circle:
\begin{equation}
y_q(\omega, t)
=\int_{\bS}
x(\bp, \omega, t)
\e^{-j\omega\la\bp, \frac{\br_q}{c}\ra}
\di\bp\quad\text{for }q=1, \cdots, Q,
\end{equation}
where $c$ is the speed of sound.
In this paper, we will take as measurements the cross-correlations\footnote{
Alternatively, we can estimate the DOA directly from the received microphone signals $y_q(\omega, t)$. We leave the detailed discussions for a future publication.
%In radio interferometry~\cite{thompson2001interferometry}, this cross-correlation is also known as \textit{visibilities} as it is related to the brightness distribution of the sky image.
} 
between the received signals for a microphone pair $(q, q')$:
\begin{equation}
V_{q, q'}(\omega)\eqdef\E{
y_q(\omega, t)
y_{q'}^*(\omega, t)
}
\label{eq:visiDef}
\end{equation}
for $q, q' \in[1, Q]$ and $q\neq q'$. In practice, $V_{q, q'}$ is estimated by averaging over frames; for simplicity we use the same symbol for the empirical version.
%\begin{definition}
%In analogy to applications in radio interferometry~\cite{thompson2001interferometry}, we refer to the cross-correlations 
%between a microphone pair $(q, q')$: 
%\begin{equation}
%V_{q, q'}(\omega)\eqdef\E{
%y_q(\omega, t)
%y_{q'}^*(\omega, t)
%}
%\label{eq:visiDef}
%\end{equation}
%as \textit{{\red visibilities}} of the sound field.
%\end{definition}

If we assume that these sources are spatially uncorrelated, 
then the cross-correlation reduces to:
{\footnotesize\begin{align}
V_{q,q'}(\omega)
%=&
%\E{
%y_q(\omega, t)
%y_{q'}^*(\omega, t)
%}
%\\
\!=&\!
\int_{\bS}\!
\int_{\bS}\!
\E{x(\bp, t)
x^*(\bp', t)} 
\!\e^{-j\omega\la\! \bp, \frac{\br_q}{c}\!\ra}
%\\
%&
\!\e^{j\omega\la\! \bp', \frac{\br_{q'}}{c}\!\ra}
\!\!\di\bp\!\di\bp'\\
=&
\int_{\bS}
I(\bp, \omega)\e^{-j\omega
\la
\bp,
\Delta\br_{q,q'}
\ra}
\di\bp\label{eq:visibilityImg},
\end{align}}%
where $\Delta\br_{q,q'} \eqdef \frac{\br_q -\br_{q'}}{c}$ is the normalized baseline.

\subsubsection{Point Sources}\label{sssec:pt_src}
Assume now that our source distribution is a sum of spatially localized sources. We can write
\begin{equation}
  \label{eq:localized_sources}
  \tilde{x}(\bp, \omega) = \sum_{k = 1}^K \alpha_k(\omega) \sqrt{\gamma}\phi\big(\gamma (\bp - \bp_k)\big)
\end{equation}
where $\phi(\bp)$ is a localized function with finite energy, $\int \phi^2\di\bp = 1$, and $\gamma$ is the spatial scaling factor. To ensure that the $k$-th source has finite power equal to $\sigma_k^2$, we ask that\linebreak
\(\int \E{|\alpha_k|^2 \gamma \phi^2(\gamma \bp)}\di\bp = \sigma_k^2\) yielding $\E{|\alpha_k|^2} = 
 \sigma_k^2$. If we now denote the intensity corresponding to \eqref{eq:localized_sources} by $I_{\gamma}(\bp, \omega)$, then in the point source limit we get
\begin{equation}
  \lim_{\gamma \to 0} I_\gamma(\bp, \omega) =: I(\bp, \omega) = \sum_{k=1}^K \sigma_k^2(\omega) \delta(\bp - \bp_k),
\end{equation}
where $\sigma_k^2(\omega)=\E{ | \tilde{x}(\bp_k, \omega) |^2 }$, $\bp_k=(\cos \varphi_k, \sin \varphi_k)\transpr$, and $\varphi_k$ is the azimuth of the $k$-th source.
%For notational brevity, we will drop the dependence on the mid-band frequency $\omega$ in the following part of this paper when there is no confusion.

With the above established, we rewrite the microphone cross-correlations~\eqref{eq:visibilityImg} as
\begin{align}
V_{q, q'}(\omega)
%&
%=\int_{\bS}
%\sum_{k=1}^K
%\sigma_k^2(\omega)
%\delta(\bp - \bp_k)
%\e^{-j\omega\la
%\bp,
%\Delta\br_{q,q'}
%\ra}
%\di\bp
%\\
&
=
\sum_{k=1}^K
\sigma_k^2(\omega)
\e^{-j\omega\la
\bp_k,
\Delta\br_{q,q'}
\ra}\label{eq:visibility}
\end{align}
for $q \neq q'$. Instead of the more conventional approach to FRI sampling where the microphone signals $y_q$ would be used as input~\cite{Hayuningtyas:2012vt}, we use as input the correlations $V_{q, q'}$. This effectively increases the number of measurements and allows us to use a small number of microphones.

\subsection{Point Source Reconstruction}
Following the generalized FRI sampling framework~\cite{pan2016towards}, we will first identify the set of unknown sinusoidal samples and its relation with the given measurements~\eqref{eq:visiDef}. Then, the DOA estimation is cast as a constrained optimization (see e.g.,~\eqref{eq:constrainedForm1}).
\subsubsection{Relation between Measurements and the Uniform Samples of Sinusoids}\label{sssec:linarMappingG}
%Spherical Representation of the Measured Visibilities
Since $\bp$ is supported on the circle, we have the following Fourier series representation for intensity:
\begin{equation}
I(\bp,\omega)
=\sum_{m\in\Z}
\hat{I}_{m}(\omega) Y_{m}(\bp),
\label{eq:ptSourceSph}
\end{equation}
where $Y_{m}(\bp)$ is the Fourier series basis $Y_m(\bp)=Y_m(\varphi)=\e^{jm\varphi}$, and $\hat{I}_{m}(\omega)$ is the associated expansion coefficient for a sub-band centered at frequency $\omega$:
{\footnotesize\begin{equation}
\hat{I}_m(\omega)=\!\frac{1}{2\pi}\!\int_{\bS}\!
I(\bp, \omega) Y_m^*(\bp) \di\bp
\!=\!\frac{1}{2\pi}\! \sum_{k=1}^K\! \sigma_k^2(\omega) \e^{-j m\varphi_k}\!.
\label{eq:uniSinusoid}
\end{equation}}%
Notice that the Fourier series coefficients $\hat{I}_m(\omega)$ for $m\in\Z$ are uniform samples of sinusoids, which are related with the cross-correlation~\eqref{eq:visibilityImg} as:
\begin{align}
&V_{q,q'}(\omega)
%\\
%&
=
\int_{\bS}
\sum_{m\in\Z}
\hat{I}_m(\omega)
Y_m(\bp)\e^{-j\omega\la \bp, \Delta\br_{q, q'}\ra}
\di\bp\\
&\stackrel{(a)}{=}
2\pi\!\sum_{m\in\Z}\!
(-j)^m
J_m\paren*{\|\omega\Delta\br_{q,q'}\|_2}
Y_{m}\paren*{\!
\frac{\Delta\br_{q,q'}
}{
\|\Delta\br_{q,q'}\|_2
}
\!}
\hat{I}_m
%\cdot\\
%&\hphantom{=\sum_{m\in\Z}\ 
%}
%\underbrace{
%\sum_{k=1}^K \sigma_k^2(\omega) \e^{-j m \varphi_k}
%}_{b_m(\omega)},
\label{eq:visi2uniSinusoid}
\end{align}
where $(a)$ is from Jacobi-Anger expansion~\cite{colton2012inverse} of the complex exponential and $J_{m}(\cdot)$ is Bessel function of the first kind.

Therefore, we establish a linear mapping from the uniformly sampled sinusoids $\hat{I}_m$ to the given measurements $V_{q,q'}$. Concretely, denote a lexicographically ordered vectorization of the cross-correlations $V_{q, q'}(\omega)$, $q \neq q'$ by $\ba(\omega) \in \C^{Q(Q-1)}$, and let the vector $\bb(\omega)$ be the Fourier series coefficients $\hat{I}_m(\omega)$ for $m \in \cal{M}$, where $\cal{M}$ is a set of considered Fourier coefficients\footnote{Note that these correspond to the spatial Fourier transform of $I$ over the circle, not to sources' temporal spectra.}. Define also a $Q(Q - 1) \times |{\cal M}|$ matrix $\GG(\omega)$ as
\begin{equation}
  g_{(q,q'), m}(\omega) \eqdef (-j)^m
J_m\paren*{\|\omega\Delta\br_{q,q'}\|_2}
Y_{m}\paren*{
\frac{\Delta\br_{q,q'}
}{
\|\Delta\br_{q,q'}\|_2
}
},
\end{equation}
where rows of $\GG$ are indexed by microphone pairs $(q, q')$, and columns of $\GG$ are indexed by Fourier bins $m$. We can then concisely write \eqref{eq:visi2uniSinusoid} as
$\ba(\omega) = \GG(\omega) \bb(\omega).$

%--------------------------------------------------------
\subsubsection{Annihilation on the Circle}
Since $\hat{I}_m$ in~\eqref{eq:uniSinusoid} is a weighted sum of uniformly sampled sinusoids, we know that $\hat{I}_m$ should satisfy a set of annihilation 
equations~\cite{vetterli2002sampling}:
$
\hat{I}_m\mathop{*} h_m = 0.
$
Here $h_m$ is the unknown annihilating filters to be recovered. 
A polynomial,
whose coefficients are specified by the filter $h_m$,
has roots located at $\e^{-j\varphi_k}$~\cite{vetterli2002sampling}.
The source azimuths $\varphi_k$ are subsequently reconstructed with polynomial root-finding.

In a multi-band setting, the uniform sinusoidal samples $\hat{I}_m(\omega)$ are different for each sub-band. This is because, the signal power $\sigma_k^2$ varies with the mid-band frequency $\omega$ in general. 
However, 
since we have \textit{the same} source locations $\varphi_k$ for each sub-band, 
we only need to find one filter $h_m$ (depending solely on the source locations $\varphi_k$) that annihilates $\hat{I}_m(\omega)$ for all $\omega$-s:
\begin{equation}
%\(\ds
\hat{I}_m(\omega) \mathop{*}_m h_m = 0\quad\forall\omega.
%\)
\end{equation}
%--------------------------------------------------------
\subsubsection{Reconstruction Algorithm}\label{sssec:reconAlg}
%We can rewrite~\eqref{eq:visibility} for any mid-band frequency $\omega\in\R$ as
%\begin{equation}
%V_{q, q'}=\sum_{k=1}^K
%\sigma_k^2(\omega)\e^{-j\la\bp_k, \omega\Delta\br_{q, q'}\ra},
%\end{equation}
%By utilising the wide-band measurements, it appears as if we had more microphones with different baselines: $\omega\Delta\br_{q, q'}$.

Following the discussion in the previous section, we reconstruct the source locations \textit{jointly} across all sub-bands.
More specifically, suppose we consider $J$ sub-bands centered around frequencies $\{\omega_j\}_{j=1}^J$. Then, we formulate the FRIDA estimate as a solution of the following constrained optimization:
%\vspace*{-0.5em}%<==use this if out of space
\begin{equation}
\begin{aligned}
\min_{
\begin{subarray}{c}
\bb_{1},\cdots,\bb_{J}\\
\bh\in\mathcal{H}
\end{subarray}
}
\hquad&
\sum_{i=1}^J
\big\|\ba_{i}-\GG_{i}\bb_{i}\big\|_2^2\\
\text{subject to}\quad & \bb_{i} * \bh = \bzero\quad\text{for }i=1,\cdots, J,
\end{aligned}
\label{eq:constrainedForm1}
\end{equation}
Here $\ba_{i}$, $\bb_{i}$ and $\GG_{i}$ are the cross-correlation, uniform sinusoidal samples, and the linear mapping between them for the $i$-th sub-band as specified in Section~\ref{sssec:linarMappingG};
$\mathcal{H}$ is a feasible set that the annihilating filter coefficients belong to, e.g., $\|\bh\|_2^2=1$. 
%%%%%%%%%%%%%%%%%%%%%%
%Further, if we concatenate $\left\{\ba^{(i)}\right\}_{i=1}^J$ and 
%$\left\{\bb^{(i)}\right\}_{i=1}^J$
%as column vectors $\ba$ and $\bb$, respectively; and denote a block diagonal matrix $\GG$ as
%\begin{equation}
%\GG=
%%\overbrace{
%\left[
%\begin{matrix}
%\GG^{(1)} & \bzero & \cdots & \bzero\\
%\bzero & \GG^{(2)} & \ddots & \vdots\\
%\vdots & \ddots & \ddots & \bzero\\
%\bzero & \cdots & \bzero & \GG^{(J)}
%\end{matrix}
%\right]
%%}^{J\text{ blocks}}
%,
%\end{equation}
%then~\eqref{eq:constrainedForm1} is equivalent to
%\begin{align}
%\min_{\bb,\bh\in\mathcal{H}}\quad&\|\ba-\GG\bb\|_2^2\\
%\text{subject to}\quad & \bb * \bh = \bzero,
%\end{align}
%%%%%%%%%%%%%%%%%%%%%%

Note that~\eqref{eq:constrainedForm1} is a simple quadratic minimization with respect to $\bb_{i}$-s for a given annihilating filter $\bh$. By substituting the solution of $\bb_{i}$ (in function of $\bh$), we end up with an optimization for $\bh$ alone:
\begin{equation}
\min_{\bh\in\mathcal{H}}\quad
\bh\transp \bm{\Lambda}(\bh) \bh,
\label{eq:opt_hOnly}
\end{equation}
where 
\vspace{-1mm}
\[
%\begin{equation}%<==use this if out of space
%\resizebox{\linewidth}{!}{$\ds
\bm{\Lambda}(\bh)\!=\!\sum_{i=1}^J
\TT\transp(\bbeta_{i})
\left[
\RR(\bh)
\paren*{
\GG_{i}\transp
\GG_{i}
}^{-1}
\RR\transp(\bh)
\right]^{-1}
\TT(\bbeta_{i}).
%$}
%\end{equation}
\] %<==use this if out of space
%Then, the uniform sinusoidal samples $\bb^{(i)}$ is:
%\begin{equation}
%\bb^{(i)}
%=\bbeta^{(i)}
%-
%\paren*{
%\big(\GG^{(i)}\big) \transp
%\GG^{(i)}
%}^{-1}
%\RR\transp(\bh)
%\end{equation}
Here $\bbeta_{i} = (\GG_{i}\transp \GG_{i})^{-1} \GG_{i}\transp\ba_{i}$; $\TT(\cdot)$ builds a Toeplitz matrix from the input vector; and $\RR(\cdot)$ is the \textit{right-dual} matrix associated with $\TT(\cdot)$ such that $\TT(\bb)\bh=\RR(\bh)\bb$, $\forall\bb,\bh$. This follows from the commutativity of convolution: $\bb*\bh = \bh*\bb$.

In general, it is challenging to solve~\eqref{eq:opt_hOnly} directly.
We use an iterative strategy, building $\Lambda(\bh)$ with the reconstructed $\bh$ from the previous iteration.
However, unlike similar approaches (e.g.~\cite{Bresler:1986hq}), we do not aim at obtaining a convergent solution of~\eqref{eq:opt_hOnly} but rather a valid solution such that 
the reconstructed sinusoidal samples $\bb_{i}$-s explain the given measurements up to a certain approximation level ($\varepsilon^2$): $\sum_{i=1}^J \|\ba_{i}-\GG_{i}\bb_{i}\|_2^2\leq\varepsilon^2$.
Readers are referred to~\cite{pan2016towards} for detailed discussions on the 
algorithmic details, e.g., choice of $\varepsilon$, implementation details, etc.

\begin{figure}[t]
  \centering
  \includegraphics[width=0.48\linewidth]{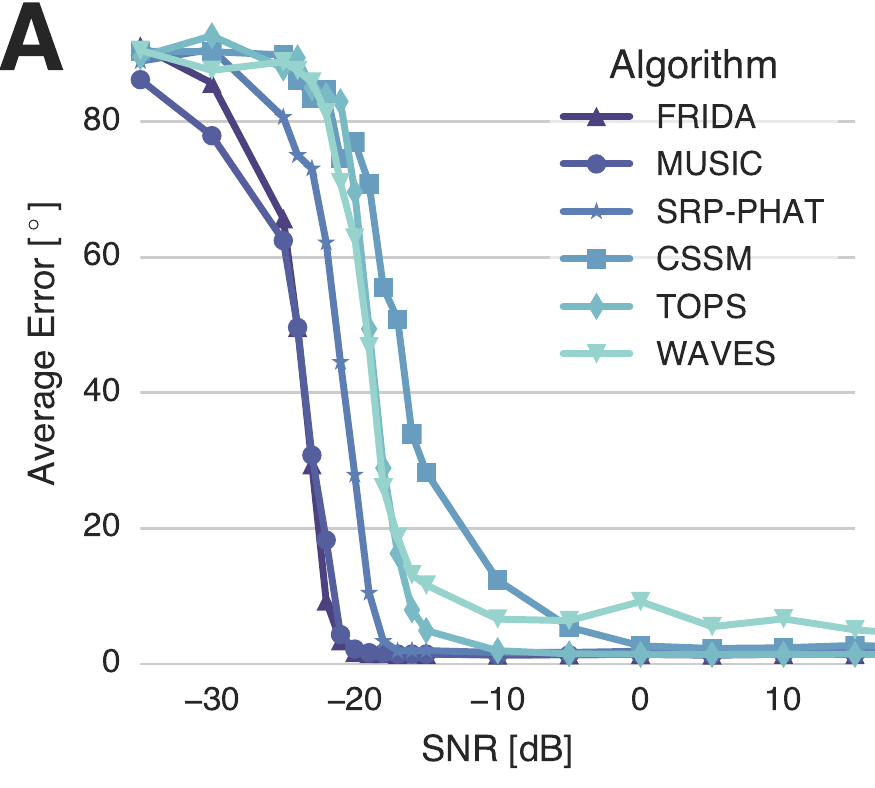}\hfill
  \includegraphics[width=0.48\linewidth]{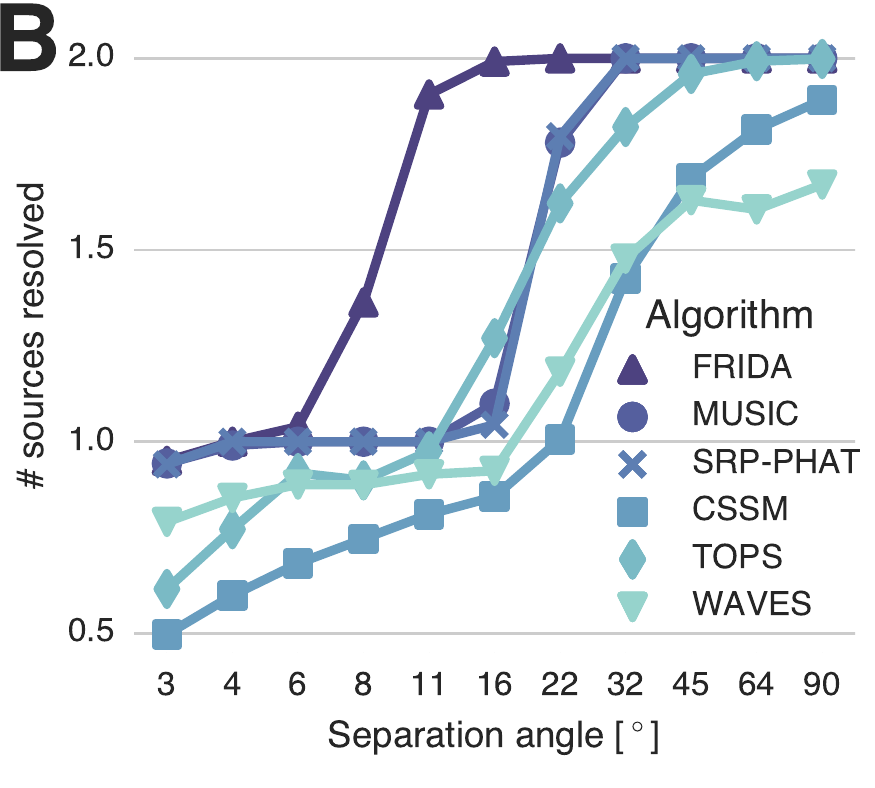}
  \caption{A Average DOA reconstruction error as a function of SNR.
    Lower is better. B Average number of sources reconstructed for the
    case of two sources separated by a fixed angle.}
  \label{fig:experiment_synthetic}
\end{figure}

\section{Experiments}

% Mean
% FRI               359.498555    4.558601
% MUSIC               1.640579  258.426788
% SRP                 1.355261  321.696057
% 1.96*std
% FRI               0.908734    0.449225
% MUSIC             0.739783  104.360093
% SRP               0.477528   26.475892

In this section, we demonstrate the effectiveness of the proposed algorithm
through numerical simulations and practical experiments.  We compare the
performance of FRIDA to that of other wideband algorithms: incoherent MUSIC
\cite{Schmidt:1986js}, SRP-PHAT \cite{DiBiase:2000uv}, CSSM \cite{Wang:1985jv},
WAVES \cite{diClaudio:2001bb}, and TOPS \cite{yoon2006tops}.

\begin{figure*}[t]
  \centering
  \includegraphics[width=\linewidth]{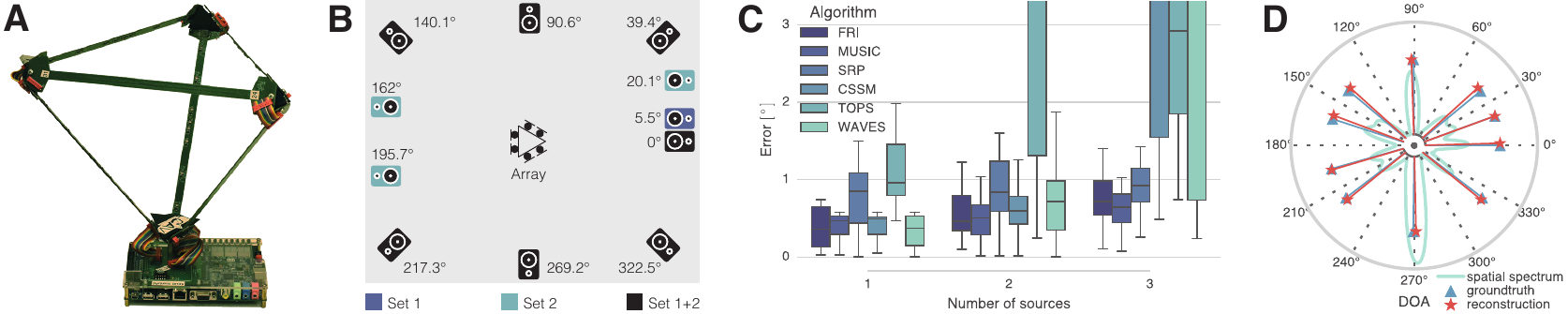} \\*[-0.5em]
  \caption{\textbf{A} Pyramic array, a compact microphone array with $48$ MEMS
    microphones distributed on the edges of a tetrahedron. For the experiments,
    only the top triangle is used. \textbf{B} Locations of the loudspeakers and
    microphone array in experiments. \textbf{C} Reconstruction error for
    the different algorithms applied to the recorded speech signals. \textbf{D}
    Reconstruction of $10$ sources from only $9$ microphones.
    The average reconstruction error is within $2^\circ$.}
  \label{fig:experiment_all}
\end{figure*}

The sampling frequency is fixed at $16$\,kHz. The narrow-band
sub-carriers are extracted by a $256$-point short-time Fourier transform (STFT)
with a Hanning window and no overlap. We use a triangular array of $24$
microphones. Each edge is $30$\,cm long and carries $8$ microphones.
%symmetrically placed, $6$ of which uniformly spaced by $4$\,cm, with an extra $2$
%microphones spaced by $8$\,mm at the center. 
The spacing of microphones ranges from $8$\,mm to $25$\,cm.
This geometry is that of the \textit{Pyramic} compact array
designed at EPFL \cite{juan_pyramic} and used to collect the recordings for the practical
experiments, see Fig.~\ref{fig:experiment_all}A. 

The number of frequency bands used (out of the $128$ narrow-bands) is a key parameter for performance and
was tuned for each algorithm. FRIDA, MUSIC and SRP-PHAT use 20
bands, CSSM and WAVES 10 bands, and TOPS 60 bands.   In the synthetic experiments, the source signals
are all white noise to simplify the choice of the sub-bands.  For speech
recordings, the STFT bins with the largest power are chosen. All implementation details are in
the supplementary material. 

The reconstruction errors are quantified according to the
distance on the unit circle defined as
\begin{equation}
  d_{\bS}(\varphi, \hat{\varphi}) = \min_{s\in \{\pm 1\}} s\,(\varphi - \hat{\varphi}) \bmod 2\pi.
  \label{eq:dist}
\end{equation}
For multiple DOA, the originals and their reconstructions are matched to 
minimize the sum of errors.{\color{white}{{\eqref{eq:dist}}}}

%$\Phi=\{\varphi_1,\dots,\varphi_K\}$, and their
%reconstruction $\hat{\Phi}$, the elements of $\Phi$ and $\hat{\Phi}$ are
%matched so as to minimize the total error, yielding the following metric
%\begin{equation*} d_{\bS}(\Phi,\hat{\Phi}) = \min_{\pi
    %\in \Pi_K} \sum_{i=1}^K d_{\bS}(\varphi_i, \hat{\varphi}_{\pi(i)}),
%\end{equation*} where $\Pi_K$ is the set of all permutations of $K$ elements.

\vspace{-1mm}

\subsection{Influence of Noise}

We study the influence of noise on the algorithms through numerical simulation.
One source playing white noise is placed at random on the unit circle. The
propagation of sound is simulated by applying fractional delay filters to
generate the microphone signals based on the array geometry. Finally, the
algorithms are run with additive white Gaussian noise of variance corresponding
to a wide range of SNR. The algorithms are fed with $256$ snapshots of $256$
samples each.  It should be noted that $256$ snapshots correspond to a
processing gain of about $24\,\dB$. We run 500 rounds of Monte-Carlo simulation
for each SNR value.

The simulation results in Fig.~\ref{fig:experiment_synthetic}A show that FRIDA
and MUSIC are the most robust with a
breaking points slightly below $-20\,\dB$.  Next are SRP-PHAT and TOPS,
breaking around $2\,\dB$ and $4\,\dB$ higher, respectively.  
While WAVES initially seems to perform as well as TOPS, it never reaches zero error.
Least resistant to
noise is CSSM, breaking down as early as $-5\,\dB$. The poor performance of WAVES
and CSSM might be attributed to poor initial estimates of the focusing
frequencies.

\begin{table}
  \centering
  \scalebox{.90}{
  \begin{tabular*}{\linewidth}{@{\extracolsep{\fill}}lccc@{}}
    \toprule
    DOA      & FRIDA      & MUSIC    & SRP-PHAT \\
    \midrule
    $0^\circ$   & $-0.5 \pm 0.4^\circ$ &    \phantom{$-9\!\!\!$}$1.6 \pm  0.3^\circ$ &   \phantom{$-3$}$1.4 \pm 0.2^\circ$ \\
    $5.5^\circ$ &  \phantom{$-$}$4.6 \pm 0.2^\circ$ &  $-93.9 \pm 41.2^\circ$ & $-38.1 \pm 8.6^\circ$ \\
    \bottomrule
  \end{tabular*}}

  \caption{The accuracy of the reconstruction for recordings with  sources
    closely located at $0^\circ$ and $5.5^\circ$. The mean is computed as the logarithm of the average
  of complex exponentials with argument given by the reconstruction angle. The second number is the
  average distance (8) from the sample to the mean.}
\label{tb:close_doa}

\end{table}

\vspace{-1mm}
\subsection{Resolving Close Sources}

Next, we study the minimum angle of separation necessary to resolve distinct
sources. We simulate two sources of white noise at angles $\varphi$ and
$\varphi+\delta$ where $\delta$ is varied from $90^\circ$ to $2.8^\circ$. The
average error is then computed over ten realizations of the noise for $120$
values of $\varphi$. We mark a DOA as successfully recovered if the
reconstruction error is less than $\delta/2$. This criterion seems crude for
large $\delta$, but for small $\delta$, where performance is critical, it is
very stringent. Here again $256$ snapshots are used and the SNR is set to $0\,\dB$.

As seen in Fig.~\ref{fig:experiment_synthetic}B, we find that FRIDA largely outperforms the other algorithms. It always
separates sources located as close as $11.2^\circ$, while the closest
contenders, MUSIC and SRP-PHAT, have difficulties for sources closer than
$22.5^\circ$. The coherent methods perform worse than the incoherent
ones; they even suffer from a lack of precision in estimating a single source.

\vspace{-1mm}
\subsection{Experiments on Recorded Signals}

Finally, we perform two experiments with recorded data to validate the
algorithm in non-ideal, real-world conditions. In the first experiment, the Pyramic
array is placed at the center of eight loudspeakers (Fig.~\ref{fig:experiment_all}B, Set 1).
All the loudspeakers are between $1.45$\,m and $2.45$\,m away from the array. Recordings are made with all possible combinations of one, two, and three
speakers playing simultaneously (distinct) speech segments of $3$ to $4$ seconds
duration. Two of the speakers are located at $5.5^\circ$ of each other to test
the resolving power of the algorithms.
%The SNR was found to be approximately $\rho \approx 37\,\dB$.

The statistics of the reconstruction errors for the different algorithms are
shown in Fig.~\ref{fig:experiment_all}C. We find the coherent methods WAVES
and CSSM to perform well for one and two sources, but break down for three
sources.  The TOPS method maintains an acceptable but somewhat imprecise
performance for more than one source. FRIDA, MUSIC and SRP-PHAT perform best
with a median error within one degree from the ground truth. Where FRIDA
distinguishes itself from the conventional methods is for closely spaced
sources. This is highlighted in Table~\ref{tb:close_doa} where the average
reconstructed DOA for the sources located at $0^\circ$ and $5.5^\circ$ is
shown. While all three methods correctly identify the first source, only FRIDA is able
to resolve the second.

The second experiment  tests the ability of FRIDA to resolve more sources than
microphones are used. We place ten loudspeakers (Fig.~\ref{fig:experiment_all}B, Set 2) around the Pyramic array and record them simultaneously
playing white noise. Then, we discard the signals of all but nine microphones
and run FRIDA. 
%The \emph{correct} reconstruction of all DOA, 
The algorithm successfully reconstructs all DOA within $2^\circ$ of
the ground truth, as shown in Fig.~\ref{fig:experiment_all}D. 
Note that none
of the subspace methods can achieve this result. While SRP-PHAT is not limited
in this way, its resolution is lower (its error is $\sim 4^\circ$ on this recording).

\vspace{-1mm}
\section{Conclusion}

We introduced FRIDA, a new algorithm for DOA estimation of sound sources. FRIDA relies on
finite rate of innovation sampling to do so efficiently on arbitrary array geometries,
avoiding any costly grid search. Its ability to use wideband signal information makes
it robust to many types of noise and interference. We demonstrate that FRIDA compares favorably
to the state-of-the-art, and clearly outperforms all other algorithms when it comes
to resolving close sources. Moreover, FRIDA is notable for resolving more sources than microphones,
as demonstrated experimentally on recorded signals.
Besides the logical extension to the full spherical case, we want to extend the 
algorithm to work on plane waves directly, rather than the cross-correlation
coefficients. This will allow the algorithm to handle correlated as well as
uncorrelated sources. Finally, it is important for practical purposes to
improve the computational complexity of the algorithm.
\bigskip

\noindent \textbf{Acknowledgement} We are indebted to Juan Azcarreta Ortiz
and Ren\'{e} Beuchat for their help with the Pyramic array.

%\begin{figure}[t]
  %\centering
  %\includegraphics[width=\linewidth]{experiment_error_box}
  %\caption{Reconstruction error for the different algorithms applied to the recorded signals.}
  %\label{fig:experiment_result}
%\end{figure}

%\begin{figure}[t]
%  \centering
%  \includegraphics[height=4cm]{experiment_9_mics_10_src}
%  \caption{Reconstruction of 10 sources using signals recorded by only 9 microphones.
%    The average reconstruction error is with $1.5^\circ$.}
%  \label{fig:experiment_result}
%\end{figure}

%
%\subsection{Synthetic Data}
%We apply the reconstruction algorithm to the received plane waves at $10$ microphones. 
%The emitted signal for the $k$-th source is a circular complex Gaussian: $\tilde{x}(\bp_k, \omega, t)\sim \mathcal{CN}(0, \sigma_k^2)$.
%Complexed-valued Gaussian white noise is added to the received signal such that the signal to noise ratio is $0\,\dB$.
%The covariance matrix~\eqref{eq:mtx_xcorr} is estimated from $256$ snapshots.
%\begin{figure}[h]
%\centering
%\includegraphics[height=0.27\textwidth]{planewave_mic0_SNR_0dB}\\
%\begin{tabular}{cc}
%\raisebox{-0.5\height}{\includegraphics[height=0.23\textwidth]{polar_numMic_10_layout30-05}}&
%\raisebox{-0.5\height}{\includegraphics[height=0.345\textwidth]{polar_K_5_numMic_10_noise_0dB_locations30-05_16_32}}
%\end{tabular}
%\end{figure}

%\nocite{*}
\bibliography{icassp2017}
\bibliographystyle{IEEEtran}

\end{document}